# Zero-field magnetometry using hyperfine-biased nitrogen-vacancy centers near diamond surfaces

Ning Wang[1,2,3], Chu-Feng Liu[1], Jing-Wei Fan[1], Xi Feng[1], Weng-Hang Leong[1], Amit Finkler[4,5,§], Andrej Denisenko[4,5], Jörg Wrachtrup[4,5], Quan Li [1,2,3 †], Ren-Bao Liu[1,2,3 †]

1. Department of Physics, The Chinese University of Hong Kong, Shatin, New Territories, Hong Kong, China

2. Centre for Quantum Coherence, The Chinese University of Hong Kong, Shatin, New Territories, Hong Kong, China

3. The Hong Kong Institute of Quantum Information Science and Technology, The Chinese University of Hong Kong, Shatin, New Territories, Hong Kong, China

4. 3rd Institute of Physics and Center for Applied Quantum Technologies, University of Stuttgart, 70569 Stuttgart, Germany

5. Max Planck Institute for Solid State Research, 70569 Stuttgart, Germany

§  Current affiliation: Department of Chemical and Biological Physics, Weizmann Institute of Science, Rehovot, Israel

**Abstract:**

Shallow nitrogen-vacancy (NV) centers in diamond are promising for nano-magnetometry for they can be placed proximate to targets. To study the intrinsic magnetic properties, zero-field magnetometry is desirable. However, for shallow NV centers under zero field, the strain near diamond surfaces would cause level anti-crossing between the spin states, leading to clock transitions whose frequencies are insensitive to magnetic signals. Furthermore, the charge noises from the surfaces would induce extra spin decoherence and hence reduce the magnetic sensitivity. Here we demonstrate that the relatively strong hyperfine coupling (130 MHz) from a first-shell $^{13}$C nuclear spin can provide an effective bias field to an NV center spin so that the clock-transition condition is broken and the charge noises are suppressed. The hyperfine bias enhances the dc magnetic sensitivity by a factor of 22 in our setup. With the charge noises suppressed by the strong hyperfine field, the ac magnetometry under zero field also reaches the limit set by decoherence due to the nuclear spin bath. In addition, the 130 MHz splitting of the NV center spin transitions allows relaxometry of magnetic noises simultaneously at two well-separated frequencies (~2.870 ± 0.065 GHz), providing (low-resolution) spectral information of high-frequency noises under zero field. The hyperfine-bias enhanced zero-field magnetometry can be combined with dynamical decoupling to enhance single-molecule magnetic resonance spectroscopy and to improve the frequency resolution in nanoscale magnetic resonance imaging.

**Main text:**

Nitrogen-vacancy (NV) centers [1–3] in diamond are promising quantum sensors of magnetic field [4–6], electric field [7], temperature [8–11], pressure [12], deformation [13], etc., for their long coherence time and correspondingly sharp spectral lines [14]. In particular, shallow NV centers in diamond are suitable for studying nano-magnetism, such as vortices in superconductors [15,16], skyrmions in magnetic thin films [17,18], spin-wave excitations in magnetic insulators [19–21], antiferromagnetism in thin films [22,23], and ferromagnetism of 2D van der Waals crystals [24]. However, the sensitivity of shallow NV centers is greatly impaired by local strains and electrical noises from surfaces [25–28], which are particularly serious under zero field, a condition often needed for measuring intrinsic magnetic properties [29–31]. Under zero field, the strain causes level anti-crossing (LAC) of the spin states, leading to clock transitions whose frequencies are invariant to first order in the magnetic field and therefore are insensitive to weak magnetic fields [32,33]. Furthermore, the symmetry breaking by the strain makes the spin transition frequency sensitive to electrical noises, which significantly reduces the spin coherence times of shallow NV centers [34]. A biasing magnetic field can shift the spin transitions away from the clock transition to suppress the effects of strain and electrical noises, but when applied to studying magnetism, it would inevitably perturb the intrinsic properties of the targets.

Here we propose and demonstrate that the relatively strong hyperfine coupling to a nearby $^{13}$C nuclear spin [Fig. 1 (a)] can effectively bias an NV center spin for enhancing zero-field magnetometry. The hyperfine coupling to a first-shell $^{13}$C (about 130 MHz) amounts to a magnetic field on the NV center electron spin about ±2.3 mT for the up/down state of the $^{13}$C nuclear spin. Considering that the strain effect on a shallow NV center spin is usually less than 10 MHz, such a hyperfine coupling can shift the spin resonances far away from the LAC, recovering the sensitivity of the spin resonance frequencies to weak magnetic signals and simultaneously suppressing the effects of electrical noises. The biasing field from the nuclear spin has no effect on sensing targets since it decays rapidly ($\sim r^{-3}$) with the distance $r$. An additional advantage of the hyperfine-biased NV magnetometry is that the relaxometry (i.e., the measurement of the spin relaxation rate $T_1^{-1}$, which equals to the noise power density at the spin transition frequency) can be simultaneously carried out at two well-separated frequencies ($\sim 2.870 \pm 0.065$ GHz) even under zero external field. The multi-frequency relaxometry can extract coarse-grained spectral information of high-frequency (~2.8 GHz) magnetic fluctuations under zero field.

The transition frequencies of the NV center electron spin ($S = 1$) are measured by optically detected magnetic resonance (ODMR) [1]. We fabricated nano-pillar arrays on a diamond membrane to increase the photon collection efficiency. One to a few shallow NV centers are created in each pillar by $^{15}$N ion implantation with a dose of $100$ ions/μm$^2$ at energy of 10 keV [35].

The Hamiltonian of the NV center electron spin and the $^{13}$C spin is

$$H = (D_{gs} + \Pi_z)S_z^2 + \Pi_x(S_y^2 - S_x^2) + \gamma_e B_z S_z + S_z \mathbf{A}_N \cdot \mathbf{I}_N + S_z \mathbf{A}_C \cdot \mathbf{I}_C, \qquad (1)$$

where $D_{gs} = 2.87$ GHz is the ground state zero-field splitting, $\Pi_{x/y/z}$ is the strain along the $x$ / $y$ / $z$ axis, $\mathbf{S}$ is the electronic spin, $\mathbf{I}_N$ is the $^{15}$N nuclear spin-1/2, $\mathbf{I}_C$ is the $^{13}$C nuclear spin-1/2, $\gamma_e$ is the gyromagnetic ratio of the electron spin, $B_z$ is the z-component of an external magnetic field, and $\mathbf{A}_N$ and $\mathbf{A}_C$ are the hyperfine coupling vectors for the nitrogen and $^{13}$C spins when the NV center spin is along the z axis. In our experiment, we keep the magnetic field along the NV axis as much as possible. Though a transverse field less than 0.1 mT cannot be fully avoided, it has little effect on our experiments. Details on the alignment of the magnetic field are given in Note 1 of Ref. [36]. Considering that the transverse magnetic field and hyperfine couplings are weak as compared with the zero-field splitting $D_{gs}$, we have dropped the non-secular terms, i.e., those linear in $S_x$ and $S_y$. We have set the $xy$ coordinates such that $\Pi_y = 0$. Under the weak magnetic condition, the Zeeman energies of the nuclear spins are neglected.

The eigen-energies of the system in Eq. (1) are [36]

$$f_{m_s,\mu_N,\mu_C} = D_{gs} + \Pi_z + m_s\sqrt{(\gamma_e B_z + \mu_N A_N + \mu_C A_C)^2 + \Pi_x^2}, \qquad (2)$$

where $m_s = 0, \pm 1$ labels different electron spin states, and $\mu_N = \pm 1/2$ and $\mu_C = \pm 1/2$ label the respective nuclear spin states. The hyperfine coupling to the host $^{15}$N is $A_N = 3.15$ MHz [37]. When there is no strongly coupled $^{13}$C spin ($A_C = 0$) and the magnetic field is small ($|\gamma_e B_z \pm A_N/2| \ll \Pi_x$), the spin resonance frequencies depend quadratically on the magnetic field due to the LAC [32,33]. When there is a $^{13}$C nuclear spin located in the first shell ($A_C \approx 130$ MHz $\gg \Pi_x$) [36,39], the eigen-energies near zero field are

$$f_{m_s,\mu_N,\mu_C} \approx D_{gs} + \Pi_z + m_s(\gamma_e B_z + \mu_N A_N + \mu_C A_C), \qquad (3)$$

depending linearly on the external field $B_z$.

For comparison, we choose two NV centers with the same crystallographic direction in two different pillars, one with no $^{13}$C spin nearby (NV1) and one with a first-shell $^{13}$C (NV2-C13) [Fig.1 (a)]. Figure 1 (b) shows the ODMR frequencies of NV1 as functions of magnetic field (see Fig. S2 in Ref. [36] for the spectra). The resonance frequencies present two branches with the LAC at $B_z = \mu_N A_N/\gamma_e \approx \mp 0.05$ mT corresponding to the $^{15}$N spin state $\mu_N = \pm 1/2$. The data are well reproduced by Eq. (2) with fitting parameters $\Pi_x = 3.16$ MHz and $\Pi_z = 3.96$ MHz. Due to the strain, the spin resonance frequencies are insensitive to the magnetic field in the range $|B_z \pm A_N/2| < \Pi_x$. The strong hyperfine coupling in NV2-C13 leads to two well-separated sets of resonances near 2815 MHz and 2940 MHz near zero field ($B < 0.2$ mT) [Fig. 1(c)], which depend linearly on the magnetic field. The LACs are shifted to $B_z = \mu_C A_C/\gamma_e \approx \mp 2.25$ mT for the two eigen states of the $^{13}$C spin [inset in Fig. 1 (c)]. The experimental data of NV2-C13 are best fitted with strains $\Pi_x = 2.20$ MHz and $\Pi_z = 1.10$ MHz. The susceptibilities ($|\partial f/\partial B_z|$) of the two NV centers near zero field [Fig. 1 (d)], derived from Eq. (2) with the strain parameters obtained from Figs. 1 (b) and (c), is largely suppressed ($\ll \gamma_e$) by the strain in NV1 but is recovered to $\gamma_e$ by the hyperfine bias of the first-shell $^{13}$C in NV2-C13.

The sensitivity of magnetometry also depends on the spin coherence time. We measure the spin coherence time $T_2^*$ using Ramsey interference (see Fig. S3 in Ref. [36] for examples). For NV1, the coherence time $T_2^*$ drops significantly at the LAC [Fig. 2 (a)]. Since at the LAC the resonance frequencies are insensitive to the variation of the magnetic field, the dephasing due to magnetic noises should have been suppressed [32–34]. Therefore, the coherence time dropping at the LAC indicates that the increased decoherence of NV1 at the LAC is mainly caused by the enhancement of the effects of electrical noises by the strain, which is consistent with literature [25–28,34] and simulation (see Note 3 in Ref. [36]). For NV2-C13, the strong hyperfine interaction shifts the LAC away from the zero field, so the coherence time is nearly a constant near zero field [Fig. 2 (b)]. The coherence time of NV2-C13 near zero field is longer than near its LAC [inset of Fig. 2 (b)], since the strong hyperfine bias suppresses the effects of electrical noises.

The dc magnetic sensitivity by Ramsey measurement depends on the susceptibility and the spin coherence time by [43]

$$\eta \approx \left|\frac{\partial f}{\partial B_z}\right|^{-1} \frac{e^{(\tau/T_2^*)^2}}{C\sqrt{P_0 t_R}} \frac{\sqrt{t_I + \tau + t_R}}{\tau}, \tag{4}$$

where $t_{I/R}$ is the spin initialization/readout time, $C$ is the signal contrast of the Ramsey interference, and $P_0$ is the photon-count rate. In our experiments, $t_I = t_R = 300$ ns and $T_2^*$ is in the order of several μs. The optimal sensitivity is taken at $\tau = T_2^*/2$ [43]. For NV1, the saturated photon-count rate $P_0 \approx 360$ kps and $C \approx 0.11$; for NV2-C13, $P_0 \approx 440$ kps and $C \approx 0.11$. As shown in Fig. 2 (c), the magnetic sensitivity of NV1 near zero field is about 20 times worse than that under a bias magnetic field. The sensitivity of NV2-C13 is nearly invariant in the small magnetic regime and is significantly better than that of NV1, owing to its longer coherence time $T_2^*$ and large susceptibilities $|\partial f/\partial B_z|$.

We demonstrate the dc sensitivities of the NV sensors using real-time magnetic field measurement. We use an electromagnet to generate small magnetic field variations. The sensitivity of the pulsed ODMR approaches that of the Ramsey sequence in Eq. (4) [43]. For fast measurement, we adopt the two-point method, in which the photon counts $C(f_i)$ are recorded in pulsed ODMR at two microwave frequencies [44], one ($f_1$) at the half-width-half maximum of the ODMR resonance and the other ($f_2$) far away from the resonance [inset of Fig. 2 (d)]. The signal $S = \frac{C(f_2) - C(f_1)}{C(f_2)}$ is measured as a function of time. The dependence of the signal on the field change $\delta B$ is $\delta S = \frac{C'(f_1)}{C(f_2)} \frac{\partial f}{\partial B_z} \delta B_z$, where $C' \equiv \frac{\partial C}{\partial f}$. Due to the strain effect, $\left|\frac{\partial f}{\partial B_z}\right| \ll \gamma_e$ for NV1 near zero field. When a biased magnetic field about 0.5 mT is applied, $\left|\frac{\partial f}{\partial B_z}\right| = \gamma_e$. As shown in Fig. 2 (d), NV1 cannot detect the variation between 14 μT and 26 μT but can detect both the

variations from 516 μT and 528 μT to 520 μT. In contrast, NV2-C13 can detect the small field variations with or without the bias field, with almost invariant signal-to-noise ratio.

As for the measurement of an ac magnetic field, the sensitivity can be improved by removing the effect of the static or slow-varying noises using spin echo or dynamical decoupling [6,45]. To evaluate the sensitivity of ac magnetometry, we measure the spin echo coherence time $T_2$ of the NV sensors near zero field. The ac magnetic sensitivity, for measurement with spin echo, is $\eta_{ac} \sim \eta_{dc}\sqrt{T_2^*/T_2}$. The coherence time $T_2$ of NV1 [Fig. 3 (a)] drops dramatically near zero field, being shorter by almost one order of magnitude than that under a bias field of 0.1 mT. The $T_2$ time of NV1 is partially recovered at the clock transition points $B_z = \mp \frac{A_N}{2\gamma_e}$ due to the suppression of the magnetic noises by the LAC. The overall drop of coherence time near zero field and the partial recovery of coherence time at the LAC suggest that both electrical and magnetic noises contribute to the echo decoherence and the former is stronger. The field-dependence of the spin coherence time of NV2-C13 is in good agreement of the nuclear spin bath theory [42], which confirms that the electrical noises are largely suppressed by the hyperfine bias. In the medium-weak field regime (0.1 mT $< B_z \ll$ 30 mT), the decoherence is mainly caused by the nuclear spin precession under the magnetic field and the hyperfine field (with the latter depending on the central spin state), and the coherence time increases with decreasing the magnetic field with dependence $\propto B_z^{-0.6}$ [42]. When the magnetic field is weaker than 0.1 mT, the nuclear spin precession frequencies would be mostly determined by the hyperfine couplings, so the coherence time saturates. These results indicate that the strong hyperfine bias can push the ac magnetometry sensitivity to the limit set by nuclear spin induced decoherence. As shown in Fig. 3 (c), the ac sensitivity of NV1 is much worse than that of NV2-C13 in the low-field regime ($<$ 0.1 mT). With increasing field ($>$ 1 mT), the sensitivities of NV1 and NV2-C13 become comparable. Until to the LAC of NV2-C13 ($B_z \approx \pm 2.3$ mT), the sensitivity of NV2-C13 is also reduced by the strain effect.

Another advantage of the strong hyperfine bias comes from the well-separated electron spin resonances under zero field. These resonances can be used for zero-field relaxometry at multiple frequencies, offering partial, although very coarse-grained, spectral information of high-frequency noises. In relaxometry, the spin relaxation time $T_1$ between two states with a transition frequency $f_i$ is related to the noise power spectrum at the frequency $S(f_i)$ by [2,3,43]

$$T_1^{-1} = \gamma_e^2 S(f_i). \qquad (5)$$

Under zero/low magnetic fields, the conventional NV relaxometry can only detect the noise at one frequency (~2.87 GHz). With the NV spin resonance split by the 130 MHz coupling of a first-shell $^{13}$C nuclear spin, the noise intensities at two frequencies (~2.870 ± 0.065 GHz) can be measured under zero field.

To demonstrate the zero-field two-frequency relaxometry, we use a shallow NV center coupled to a first-shell $^{13}$C (NV3-C13) to measure the magnetic noise from a magnetic nanoparticle (MNP). The spin relaxation rate of this NV center without the MNP is 0.15(1) ms$^{-1}$, approaching the typical values in bulk diamond. We dropcast the ethanol solution containing copper-nickel MNPs on the diamond surface. The presence of a nearby MNP is evidenced by the splitting of the NV resonances [Fig. 4 (a)]. The spin relaxation times are significantly shorter than before the introduction of the MNP. The different relaxation dynamics of the states $|m_s = 0, \mu_C = -1/2\rangle$, $|m_s = -1, \mu_C = +1/2\rangle$, and $|m_s = -1, \mu_C = -1/2\rangle$ [Fig. 4 (b)] result from the unequal relaxation rates between the states [36]. Solving the rate equation [36], we obtain the transition rate between $|m_s = 0, \mu_C = -1/2\rangle$ and $|m_s = -1, \mu_C = -1/2\rangle$ is 5.0(4) ms$^{-1}$ while that between $|m_s = 0, \mu_C = +1/2\rangle$ and $|m_s = -1, \mu_C = +1/2\rangle$ is 8.9(3) ms$^{-1}$, larger than the former one by 78%. Similarly, the transition rate between $|m_s = 0, \mu_C = +1/2\rangle$ and $|m_s = +1, \mu_C = +1/2\rangle$ is 6.0(4) ms$^{-1}$ while that between $|m_s = 0, \mu_C = -1/2\rangle$ and $|m_s = +1, \mu_C = -1/2\rangle$ is 10.0(5) ms$^{-1}$ [36], larger than the former by 67%. These differences suggest that the noise from the MNP has a sub-GHz spectral width. Such spectral information, though with a low resolution and at only two frequencies, provides valuable information for understanding the intrinsic dynamics of the MNP (without the interference from an external field).

To conclude, we have proposed a scheme to utilize strongly coupled $^{13}$C nuclear spins to bias the NV center spin resonances, which can suppress the effect of electrical noises from surfaces and to remove the insensitivity of the resonance frequencies near the level anti-crossing due to strain under zero or very weak field. The sensitivity of both dc and ac magnetometry under zero field is significantly improved by the strong hyperfine bias. The sensitivity at the level-crossing points can be further improved by using a circular polarized microwave to selectively excite the spin resonance $m_s = +1$ or $m_s = -1$ [46]. The strong coupling to the nuclear spin enables relaxometry at well-separated frequencies under weak field, which can provide partial spectral information of the noise. It is possible to have more than one first-shell $^{13}$C spins (with a small probability, e.g., ~0.03% in a natural-abundance diamond for two $^{13}$C spins in the first shell, which can be increased by isotope enrichment). Multiple $^{13}$C spins (or some other dark spins such as P1 center spins) can enable noise spectroscopy at more frequencies (e.g., 2.87 ± 0.065 GHz and 2.87 ± 0.130 GHz for two $^{13}$C spins). The hyperfine-biased NV magnetometry near zero field will be useful for the study of critical magnetic fluctuations of nanosystems, the magnetic dynamics of soft magnetic materials, and magnetic resonance imaging of delicate textures, all of which require minimum disturbance from external magnetic field.

## Acknowledgements

This work was supported by Hong Kong Research Grants Council Collaborative Research Fund Project C4007-19G, Hong Kong Research Grans Council - French ANR Joint Scheme Project A-CUHK404/18, and The Chinese University of Hong Kong Group Research Grants. The authors acknowledge the discussions with V. Jacques on the effects of noises on shallow NV centers.

[†]Corresponding authors: liquan@phy.cuhk.edu.hk, rbliu@cuhk.edu.hk

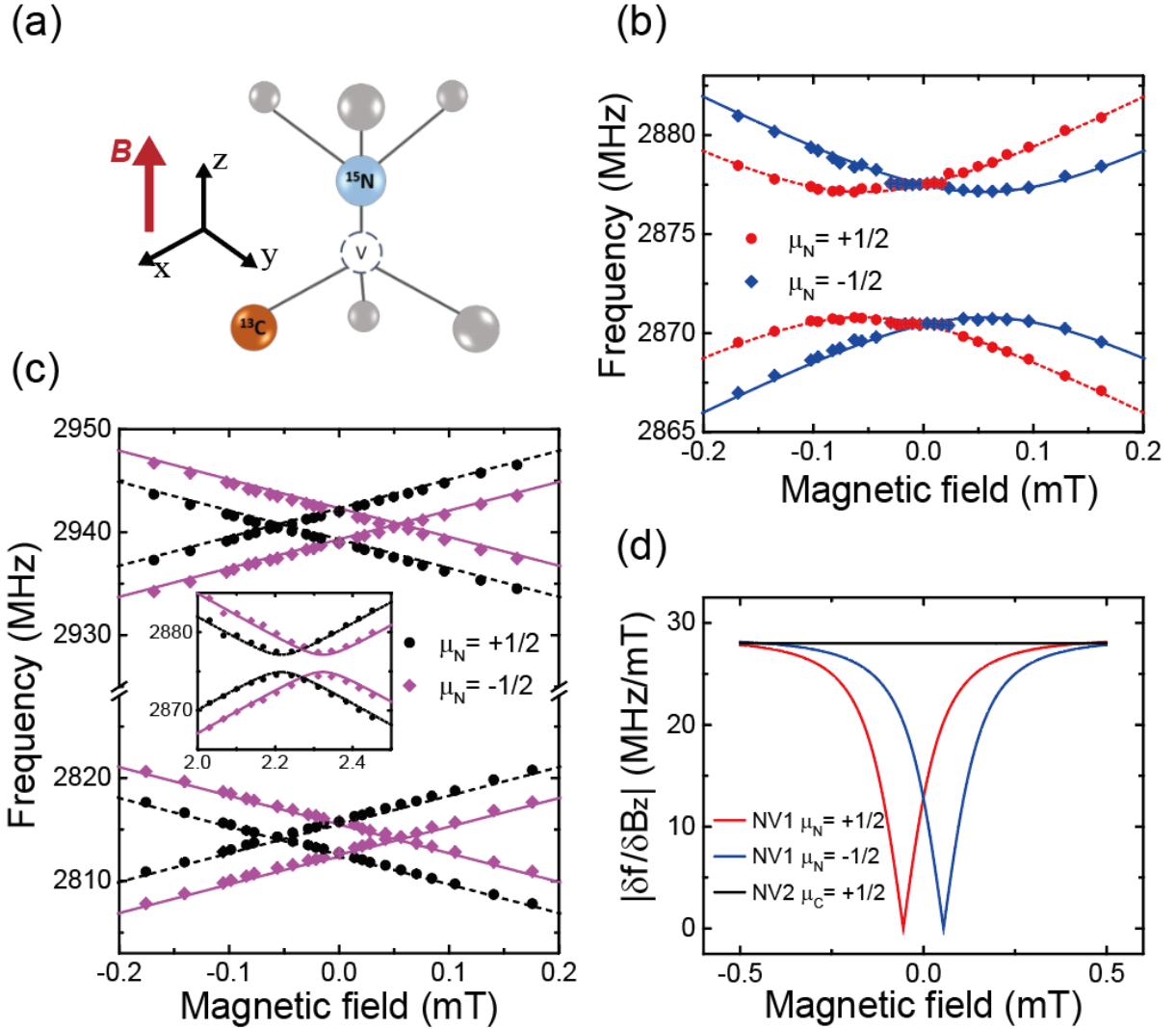

**Fig. 1. Hyperfine spin resonances of NV centers**. (a) Structure of an NV center with a host $^{15}$N nuclear spin and a first-shell $^{13}$C nuclear spin. The external magnetic field **B** is applied along the NV axis (the $z$ axis) and the $x$ axis is defined by the strain. (b) Spin resonance frequencies of an NV center without strong coupled $^{13}$C (NV1) as functions of the magnetic field. The red dots and blue diamonds are for the $^{15}$N nuclear spin states with $\mu_N = +1/2$ and $-1/2$, respectively. The corresponding lines are calculated results with strains $\Pi_x = 3.16$ MHz and $\Pi_z = 3.96$ MHz. (c) Spin resonance frequencies of an NV center with a first-shell $^{13}$C (NV2-C13). The black dots and purple diamonds are for the $^{15}$N nuclear spin states $\mu_N = +1/2$ and $-1/2$, respectively. The inset shows results at fields close to the LAC at $B_z \approx 2.3$ mT. The corresponding lines are calculated results with $\Pi_x = 2.20$ MHz and $\Pi_z = 1.10$ MHz. (d) Susceptibilities of spin

resonance frequencies to the magnetic field, $|\partial f/\partial B_z|$, near zero magnetic field for NV1 (red and blue) and NV2-C13 (black).

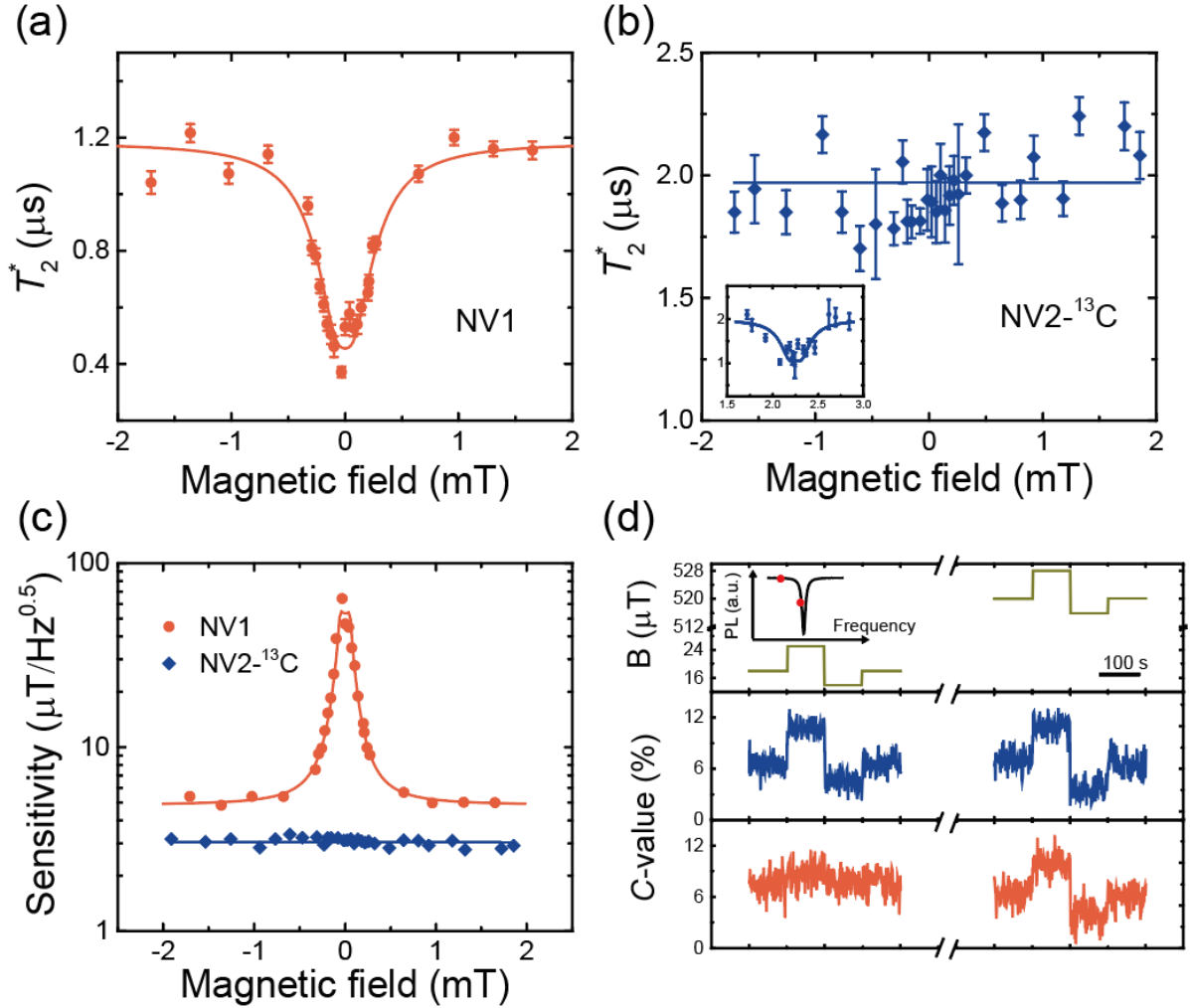

**Fig. 2. Zero-field dc magnetometry by an NV center without or with a first-shell $^{13}$C.** (a) and (b) Spin coherence time ($T_2^*$) of NV1 and NV2-C13, respectively. The dots are measured data and the lines are fitting curves. (c) Sensitivity of NV1 (orange) and NV2-C13 (blue), estimated with Eq. (4) using the parameters extracted from the experimental data. (d) Real-time magnetic field measurement using NV1 and NV2-C13. The two microwave frequencies $f_{1/2}$ are indicated by red dots in the inset of the upper panel. The acquisition time is 0.2 s per frequency. An electromagnet is used to generate magnetic field as shown in the upper panel. The middle and lower panels show the measured results by NV2-C13 and NV1, respectively.

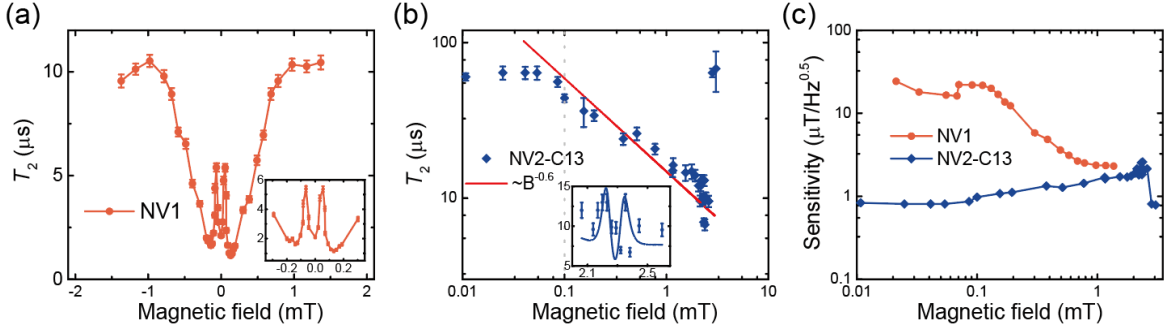

**Fig. 3. Echo coherence time and sensitivity of ac magnetometry by an NV center without or with a first-shell $^{13}$C.** (a) and (b) Spin coherence time ($T_2$) of NV1 and NV2-C13, respectively. Inset in (a) zooms in the field dependence near the LAC. (c) Sensitivities of NV1 (orange) and NV2-C13 (blue) using the parameters extracted from the experimental data as functions of magnetic field.

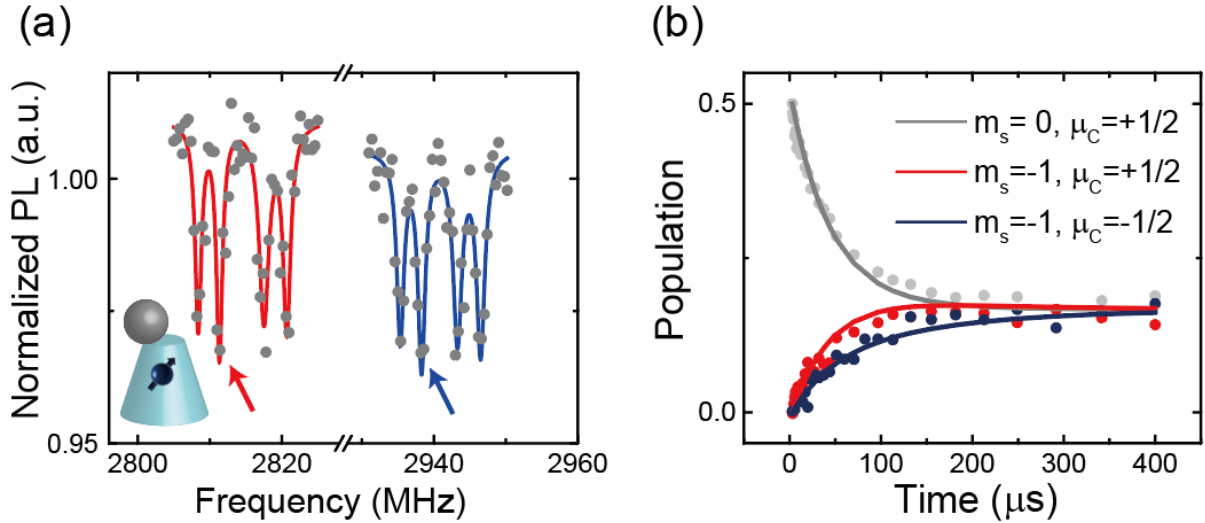

**Fig. 4. Two-frequency zero-field relaxometry enabled by hyperfine coupling to a first-shell $^{13}$C nuclear spin.** (a) ODMR spectrum of NV3-C13 with a magnetic nanoparticle placed in the proximity. (b) Populations in the states $|m_s = -1, \mu_C = +1/2\rangle$ (red), $|m_s = -1, \mu_C = -1/2\rangle$ (navy) and $|m_s = 0, \mu_C = +1/2\rangle$ (gray) as functions of time after the initialization.

# Supplementary Material

## for

## Zero-field magnetometry using hyperfine-biased nitrogen-vacancy centers near diamond surfaces

Ning Wang[1,2,3], Chu-Feng Liu[1], Jing-Wei Fan[1], Xi Feng[1], Weng-Hang Leong[1], Amit Finkler[4,5,§],, Andrej Denisenko[4,5], Jörg Wrachtrup[4,5], Quan Li [1,2,3 †], Ren-Bao Liu[1,2,3 †]

6. Department of Physics, The Chinese University of Hong Kong, Shatin, New Territories, Hong Kong, China
7. Centre for Quantum Coherence, The Chinese University of Hong Kong, Shatin, New Territories, Hong Kong, China
8. The Hong Kong Institute of Quantum Information Science and Technology, The Chinese University of Hong Kong, Shatin, New Territories, Hong Kong, China
9. 3rd Institute of Physics and Center for Applied Quantum Technologies, University of Stuttgart, 70569 Stuttgart, Germany
10. Max Planck Institute for Solid State Research, 70569 Stuttgart, Germany

§ Current affiliation: Department of Chemical and Biological Physics, Weizmann Institute of Science, Rehovot, Israel

# Contents

# Note 1. ODMR setup

The optically detected magnetic resonance (ODMR) setup is built on a commercial microscope frame. The continuous-wave laser beam from a solid-state laser (MGL-III-532-200 mW, CNI) is gated by an acousto-optic modulator and guided to the microscope frame by a single-mode fiber. A 60X oil objective is used to focus the laser beam to the samples and to collect the fluorescence. The fluorescence is filtered by a 650 nm long-pass filter and then measured by APDs (SPCM-AQRH-15-FC, Excelitas). The fiber cores of the APDs also act as the pinhole for the confocal imaging. A 3-axis piezo stage is used to scan the sample. Microwave from a signal generator (R&S SMIQ03B) is gated by a RF switch, amplified (ZHL-16W-43+, Mini-Circuits) and then fed to the PCB which holds the samples. We use the one-step e-beam evaporation process to fabricate an Omega antenna (100 nm/500 nm Cr/Au) on the cover glass with a designed metal mask. The antenna with a radius of about 80 µm is used to deliver microwave pulses to the NV centers.

The external magnetic field is generated by an electromagnet driven by a precise current source (TED4015, Thorlabs). To minimize the transverse magnetic field in our system, we first set the magnetic field at around 4 mT and finely tuned the orientation of the electromagnet to align the magnetic field along the NV axis as much as possible. This is achieved by minimizing the ground-state splitting of an NV center spin coupled to a first-shell $^{13}$C nuclear spin (NV2-C13) (see Supplementary Note 4.1 for the ground-state splitting induced by a transverse field), such that the splitting is less than the half width of the spin resonances. Once the magnetic field is aligned, we keep the orientation of the magnetic field and decrease the current of the TED4015 to reduce the magnetic field. For all fields in the measurement, the ground-state spin splitting is negligible, which means the transverse magnetic field in our system is less than 0.1 mT. Such a small transverse magnetic field (less than 0.1 mT) has negligible effects on our measurements. The magnetic field along the z-axis is calibrated by measuring the ODMR splitting of an NV center with a first-shell $^{13}$C (NV2-C13) at different currents, which shows a linear dependence. From the linear curve fitting, we obtain the magnetic field strength at a given current. Note that the earth magnetic field along the $z$-axis is compensated.

# Note 2. Hamiltonian of the system

## Note 2.1. Hamiltonian of NV electron spin and $^{15}$N nuclear spin

The Hamiltonian of NV center is

$$H = (D_{gs} + \Pi_z)S_z^2 + \Pi_x(S_yS_y - S_xS_x) + \Pi_y(S_xS_y + S_yS_x) + \gamma_e \mathbf{B} \cdot \mathbf{S} + \mathbf{S} \cdot \mathbb{A}_N \cdot \mathbf{I}_N, \quad (S1)$$

where $S_{x/y/z}$ is the spin-1 operator of the central electron spin along the $x/y/z$-direction, $\mathbf{I}_N$ is the spin-1/2 of the host $^{15}$N nuclear spin, $\Pi_{x/y/z}$ is the strain along the corresponding axis, and $\mathbb{A}_N$ is the hyperfine tensor between the central electron spin and the $^{15}$N nuclear spin, which has only diagonal elements [37] as $A_{N,xx} = A_{N,yy} = 3.65$ MHz, $A_{N,zz} = 3.15$ MHz. We can set the $xy$-axes such that $\Pi_y = 0$. Near zero field, as the transverse magnetic field $\gamma_e B_\perp \ll D_{gs}$, we only consider the magnetic field along z-axis (see Note 4.1 for discussion on the effects of a transverse field). For the same reason, we also neglect the transverse hyperfine coupling $A_{N,xx}$ and $A_{N,yy}$. The Hamiltonian of the system becomes

$$H \approx (D_{gs} + \Pi_z)S_z^2 + \Pi_x(S_yS_y - S_xS_x) + \gamma_e B_z S_z + A_{N,zz} S_z I_{N,z}. \quad (S2)$$

The eigenvalues of the Hamiltonian are

$$f_{m_s,\mu_N} = D_{gs} + \Pi_z + m_s \sqrt{(\gamma_e B + \mu_N A_N)^2 + \Pi_x^2}, \quad (S3)$$

Here $m_s = 0, \pm 1$ labels different states of the electron spin, $\mu_N = \pm 1/2$ are eigenvalues of the nuclear spin along the z axis and $A_N = A_{N,zz} = 3.15$ MHz is the hyperfine interaction along the z-axis.

## Note 2.2. Hamiltonian of NV electron spin and $^{15}$N and $^{13}$C nuclear spins

Then we consider the coupling to a first-shell $^{13}$C nuclear spin $\mathbf{I}_C$. The Hamiltonian is

$$H = (D_{gs} + \Pi_z)S_z^2 + \Pi_x(S_yS_y - S_xS_x) + \gamma_e B_z S_z + \mathbf{S} \cdot \mathbb{A}_N \cdot \mathbf{I}_N + \mathbf{S} \cdot \mathbb{A}_C \cdot \mathbf{I}_C. \quad (S4)$$

The hyperfine tensor $\mathbb{A}_C$ for a first-shell $^{13}$C spin is [38, 39]

$$\mathbb{A}_C = \begin{bmatrix} A_{C,x'x'} & 0 & A_{C,x'z} \\ 0 & A_{C,y'y'} & 0 \\ A_{C,zx'} & 0 & A_{C,zz} \end{bmatrix}, \quad (S5)$$

with $A_{C,x'x'} = 189.3$ MHz, $A_{C,y'y'} = 128.4$ MHz, $A_{C,zz} = 128.0$ MHz and $A_{C,x'z/zx'} = 24.1$ MHz [39]. Note that the $x'$-axis (set in the plane of the NV axis and the displacement of the $^{13}$C) is not necessarily the same as the $x$-axis defined by the strain.

Under the secular approximation (with the terms including $S_{x/y}$ dropped), the Hamiltonian in Eq. (1) in the main text is obtained as

$$H = (D_{gs} + \Pi_z)S_z^2 + \Pi_x(S_yS_y - S_xS_x) + \gamma_e B_z S_z + S_z \mathbf{A}_N \cdot \mathbf{I}_N + S_z \mathbf{A}_C \cdot \mathbf{I}_C, \quad (S6)$$

where $\mathbf{A}_N = A_{N,zz}\mathbf{e}_z$ and $\mathbf{A}_C = A_{C,zx'}\mathbf{e}_{x'} + A_{C,zz}\mathbf{e}_z$ are the hyperfine coupling vectors for the nitrogen and $^{13}$C spins when the NV center spin is along the $z$ axis. The eigen energies of the Hamiltonian are

$$f_{m_s,\mu_N,\mu_C} = D_{gs} + \Pi_z + m_s\sqrt{(\gamma_e B + \mu_N A_N + \mu_C A_C)^2 + \Pi_x^2}, \tag{S7}$$

with $A_C = \sqrt{A_{C,zx'}^2 + A_{C,zz}^2} \approx 130$ MHz and $A_N = 3.15$ MHz.

In our experiment, NV1 and NV2-C13 have the same orientation, which is confirmed by their nearly identical ODMR splitting under $B = 33.2$ mT as illustrated in Fig. S1.

Figure S2 shows ODMR spectra of NV1 and NV2-C13 under various weak magnetic fields.

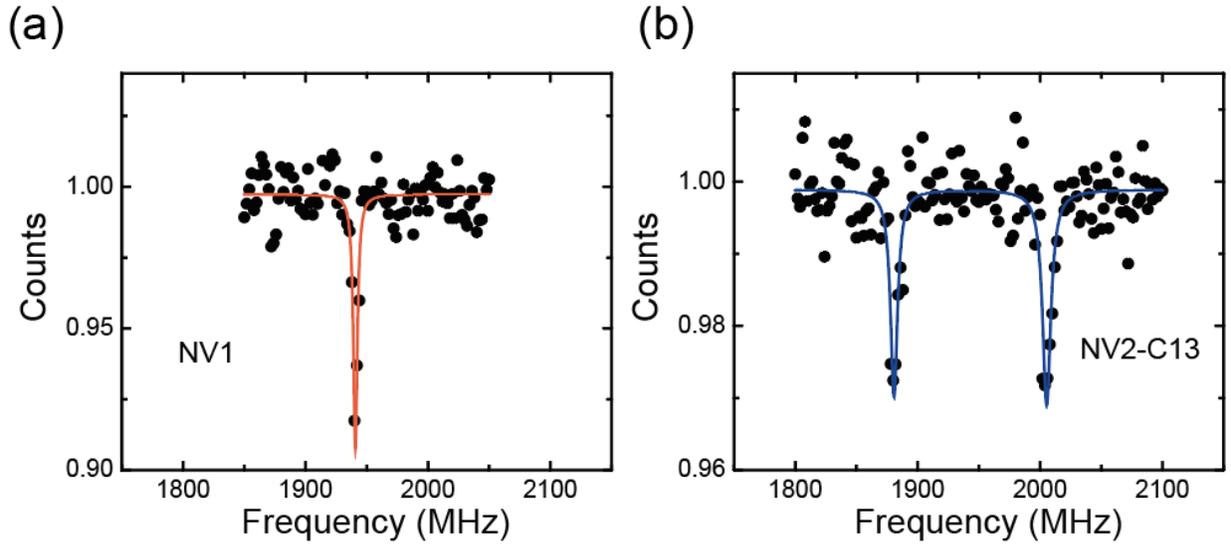

**Figure S1**. ODMR spectra of NV1 and NV2-C13 under $B = 33.2$ mT.

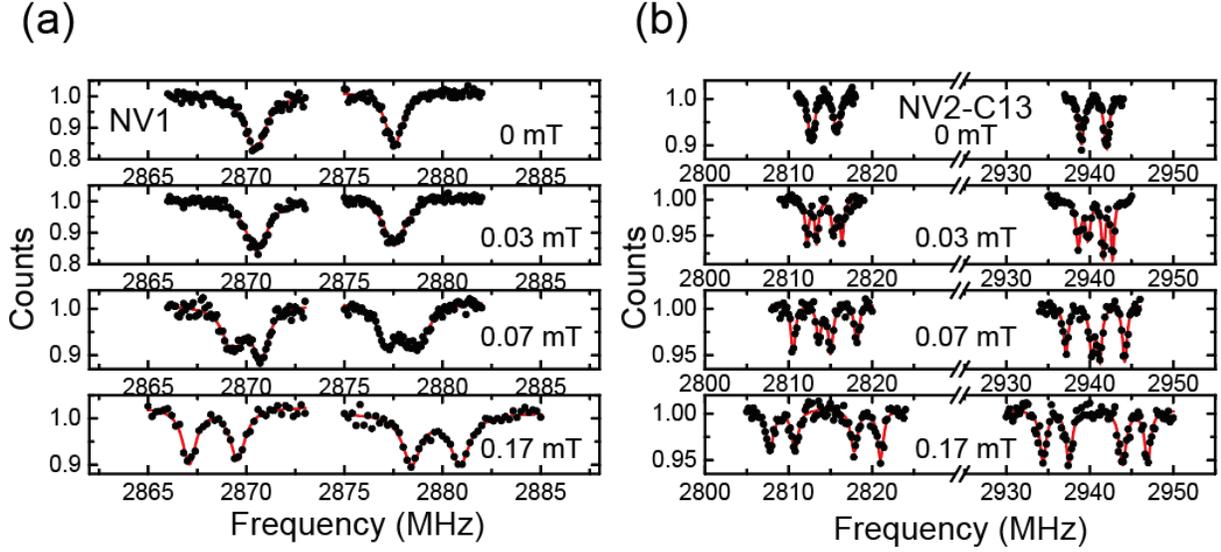

**Figure S2.** ODMR spectra of NV1 and NV2-C13 under various weak magnetic fields.

## Note 3. Decay time of Ramsey interference

The coherence time $T_2^*$ near the clock transition is measured by Ramsey interference. We denote the electrical fluctuations along different axes as $\epsilon_{x/y/z}$ and magnetic noise along the z axis as $b$. Considering the transition between the states with $m_s = 0$ and $m_s = +1$, e.g., the resonance frequency is

$$\omega_\mu = D_{gs} + \Pi_z + \epsilon_z + \sqrt{(B_\mu + b)^2 + (\Pi_x + \epsilon_x)^2 + \epsilon_y^2}, \tag{S8}$$

where $B_\mu = \gamma_e B_z + \mu_N A_N/2$ for NV1 and $B_\mu = \gamma_e B_z + \mu_N A_N/2 + \mu_C A_C/2$ for NV2-C13. The Ramsey coherence of this transition is

$$L(t) = \langle e^{-i\omega_\mu t} \rangle, \tag{S9}$$

which is averaged over the noises. We assume that both the magnetic noise and the electrical noise are Gaussian, with respective widths $\sigma_b$ and $\sigma_{x/y/z}$. Since usually the strength of the electric field coupling to the z axis is much smaller than that coupling to the x axis or y axis ($\epsilon_z \ll \epsilon_{x/y}$) [7], in the following we drop $\epsilon_z$.

By Taylor expansion up to the leading order of the noises (which are much smaller than $\Pi_x$), we get

$$\omega_\mu \approx \Delta_\mu + \frac{B_\mu b}{\Delta_\mu} + \frac{\epsilon_x \Pi_x}{\Delta_\mu}, \tag{S10}$$

with $\Delta_\mu = \sqrt{B_\mu + \Pi_x^2}$. After Gaussian integration over the noise distribution, we obtain

$$L(t) \approx \exp\left(-\frac{\sigma_x^2 \Pi_x^2 + \sigma_b^2 B_\mu^2}{2\Delta_\mu^2} t^2\right) \equiv e^{-\left(\frac{t}{T_2^*}\right)^2}. \tag{S11}$$

For fields far away from the level anti-crossing (LAC), $B_\mu \gg \Pi_x$ and $\Delta_\mu \approx B_\mu$, the electrical noise can be neglected, and therefore $L(t) \propto e^{-\frac{\sigma_b^2 t^2}{2}}$, with the coherence time independent of the magnetic field. Near the LAC (NV1 at zero or low field or NV2-C13 at a field $B \approx \pm\frac{A_C}{2\gamma_e} \approx \pm 2.3$ mT), $B_\mu \sim \Pi_x$, and the electrical noise becomes important.

We fit the measured Ramsey signal using Eq. (S11) to obtain $T_2^*$ under different magnetic fields. The Ramsey signals under different magnetic fields for NV1 and NV2-C13 are shown in Fig. S3.

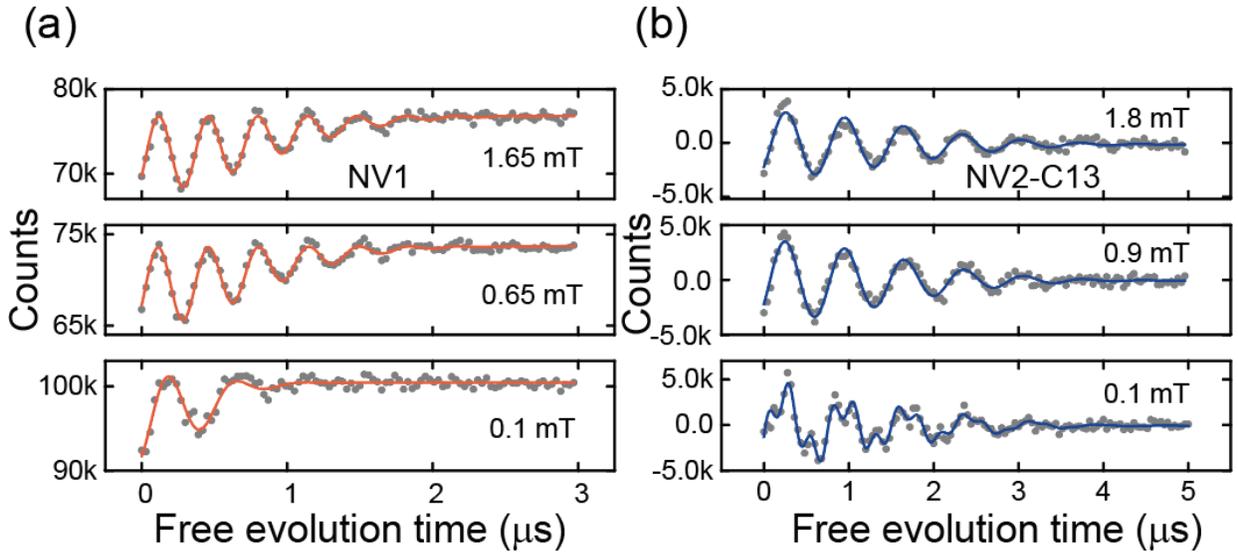

**Figure S3.** Ramsey interference signals of NV1 and NV-2C13 under different magnetic fields.

## Note 4. Spin echo measurement

### Note 4.1 Modulation induced by a first-shell $^{13}$C nuclear spin

The non-secular coupling between the NV center spin and a first-shell $^{13}$C nuclear spin can cause the mixing of the spin states and hence electron spin echo modulation (ESEEM). To simplify the discussion, we neglect the $^{15}$N nuclear spin. The Hamiltonian is

$$H = (D_{\text{gs}} + \Pi_z)S_z^2 + \Pi_x(S_y S_y - S_x S_x) + \gamma_e \mathbf{B} \cdot \mathbf{S} + \mathbf{S} \cdot \mathbb{A}_C \cdot \mathbf{I}_C. \tag{S12}$$

Here we have dropped the nuclear spin Zeeman energy for the field being small. Due to the strong hyperfine interaction, the strain effects can be neglected. The above Hamiltonian has the secular terms $H_0$ and the non-secular terms $V$ as

$$H = H_0 + V, \tag{S13a}$$

$$H_0 = D_{gs}S_z^2 + \gamma_e B_z S_z + A_{C,zz}S_z I_{C,z} + A_{C,zx'}S_z I_{C,x'}, \tag{S13b}$$

$$V = A_{C,x'x'}S_{x'}I_{C,x'} + A_{C,y'y'}S_{y'}I_{C,y'} + A_{C,x'z}S_{x'}I_{C,z} + \gamma_e B_{x'}S_{x'} + \gamma_e B_{y'}S_{y'}. \tag{S13c}$$

The exact eigen states of the secular part of the Hamiltonian ($H_0$) in Eq. (S13a) are $|m_s, \mu_C = \pm 1/2\rangle$, with eigen energies

$$E_{0,\mu_C}^0 = 0; \tag{S14a}$$

$$E_{m_s(\neq 0),\mu_C}^0 = D_{gs} + m_s\gamma_e B_z + \mu_C\sqrt{A_{C,zx'}^2 + A_{C,zz}^2}. \tag{S14b}$$

Here the quantization axis of the nuclear spin is along the direction of the hyperfine interaction, i.e., the direction of $\mathbf{e}_z \cdot \mathbb{A}_C$. The non-secular hyperfine coupling and the transverse magnetic field induce the mixing between ground states $|0, +1/2\rangle$ and $|0, -1/2\rangle$, as illustrated in Fig. S4 (a). By second-order degenerate perturbation theory, the effective Hamiltonian in the subspace expanded by the states $|0, \pm 1/2\rangle$ is

$$\begin{aligned}H_{\text{eff}}^{(2)} &= -\frac{1}{D_{gs}}\sum_{m_s=\pm 1,\mu=\pm\frac{1}{2}}\begin{pmatrix} V_{0,+1/2;m_s,\mu}V_{m_s,\mu;0,+1/2} & V_{0,+1/2;m_s,\mu}V_{m_s,\mu;0,-1/2} \\ V_{0,-1/2;m_s,\mu}V_{m_s,\mu;0,+1/2} & V_{0,-1/2;m_s,\mu}V_{m_s,\mu;0,-1/2} \end{pmatrix} \\ &= -\begin{pmatrix} \frac{A_{C,ns}^2 + 4\gamma_e(B_{x'}^2+B_{y'}^2)+4\gamma_e B_{x'}A_{C,x'z}}{4D_{gs}} & \frac{\gamma_e B_{x'}A_{C,x'x'}-i\gamma_e B_{y'}A_{C,y'y'}}{D_{gs}} \\ \frac{\gamma_e B_{x'}A_{C,x'x'}+i\gamma_e B_{y'}A_{C,y'y'}}{D_{gs}} & \frac{A_{C,ns}^2 + 4\gamma_e(B_{x'}^2+B_{y'}^2)-4\gamma_e B_{x'}A_{C,x'z}}{4D_{gs}} \end{pmatrix},\end{aligned} \tag{S15}$$

where $A_{C,ns}^2 = A_{C,x'x'}^2 + A_{C,x'z}^2 + A_{C,y'y'}^2$. The eigen-energies of the two ground states, denoted as $|m_s = 0, \pm\rangle$, are

$$E_{0,\pm} \approx -\frac{A_{C,ns}^2 + 4\gamma_e^2 B^2 \sin^2\theta \pm 4\gamma_e B \sin\theta\sqrt{(A_{C,x'x'}^2 + A_{C,x'z}^2)\cos^2\phi + A_{C,y'y'}^2\sin^2\phi}}{4D_{gs}}, \tag{S16}$$

where $\theta$ and $\phi$ are the polar and azimuthal angles of the magnetic field in the $(x', y', z)$ coordinate system, as illustrated in Fig. S4 (c). The energy difference between the two nuclear spin states is [38, 39]

$$\delta \approx \frac{2\gamma_e B_0 \sin\theta}{D_{gs}}\left(\sqrt{A_{C,x'x'}^2 + A_{C,x'z}^2}\cos^2\phi + A_{C,y'y'}\sin^2\phi\right). \tag{S17}$$

Since the nuclear spin states are quantized along different directions for $m_s = 0$ and $m_s = \pm 1$, the transitions between $|m_s = 0, \pm\rangle$ and the upper states $|m_s = \pm 1, \mu_C = \pm 1/2\rangle$ are all allowed. Therefore, a microwave pulse with Rabi frequency $\Omega > \delta$ can couple the two nuclear spin states $|m_s = 0, \pm\rangle$ to an upper state, say, $|m_s = -1, +1/2\rangle$ [see Fig. S4 (b)] and generate a coherent superposition of $|m_s = 0, +\rangle$ and $|m_s = 0, -\rangle$. Such a superposition oscillates with frequency $\delta$, inducing ESEEM at this frequency. This ESEEM can be employed to sensitively measure the splitting $\delta$, providing a sensitive magnetometry of weak transverse field. It is noted that the host $^{15}$N nuclear spin will further split the $|\pm 1, \mu_C\rangle$ state with $\Delta = 3.15$ MHz. When the Rabi frequency $\Omega > \Delta$, the ESEEM with frequency $\Delta$ is also observed.

The solid lines in Fig. S4 (d) give the energies of the two ground states ($|0, \pm\rangle$) and their energy splitting $\delta$ as functions of the transverse magnetic field by directly solving the full Hamiltonian in Eq. (S12) with strains $\Pi_z = 1.1$ MHz and $\Pi_x = 2.2$ MHz. The orientation of the magnetic field is fixed with $\theta = \pi/3$ and $\phi = \pi/6$. The corresponding dashed lines are results from the analytical expression in Eq. (S16) and (S17). The energy splitting $\delta$ of the ground state is sensitive to the transverse magnetic field ($B\sin\theta$) and the effective gyromagnetic ratio is about $0.16\gamma_e$ from the numerical simulation, a two-orders-of-magnitude enhancement of the gyromagnetic ratio of the nuclear spin.

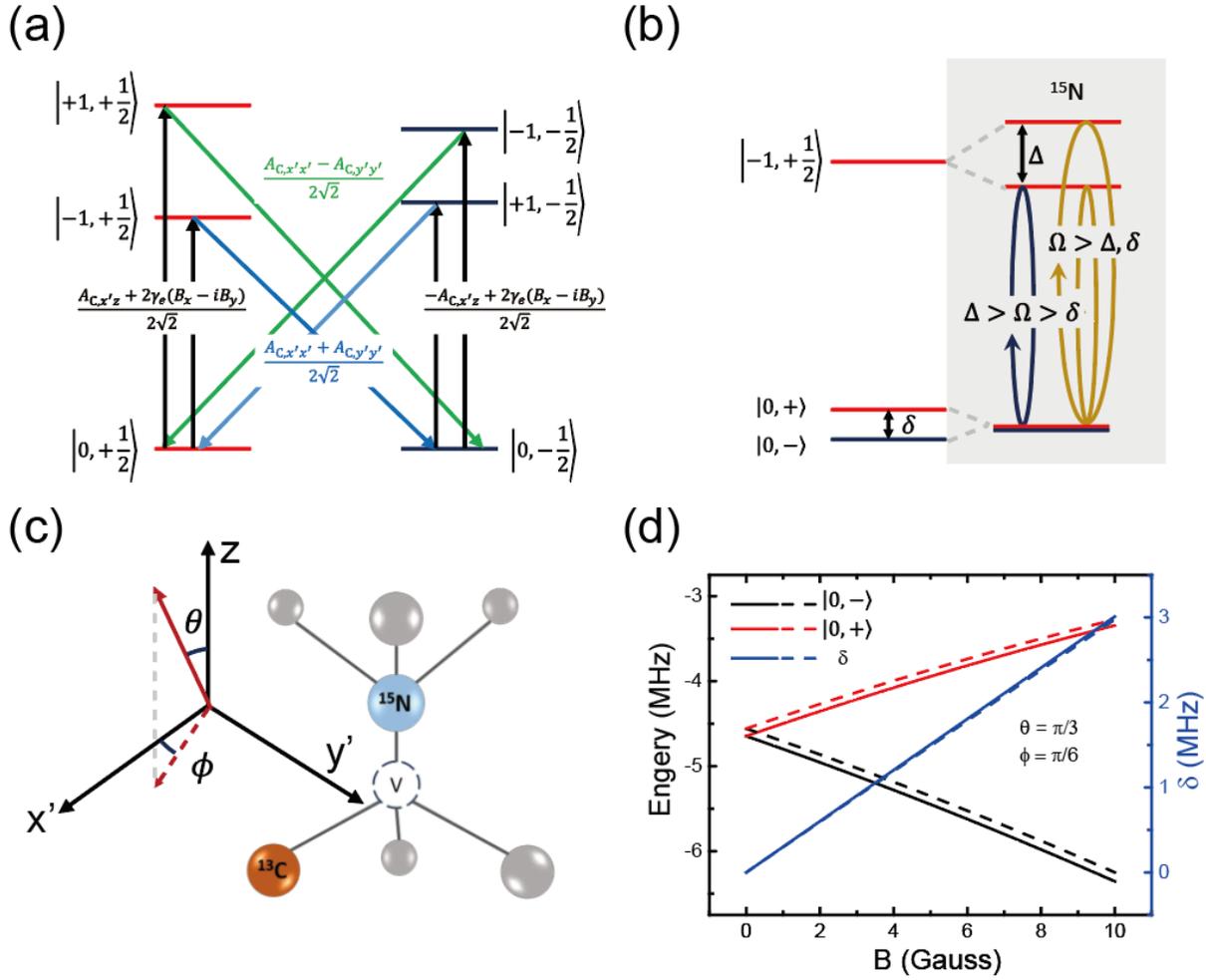

**Figure S4.** (a) Energy-level diagram for the mixing of the ground states. (b) Energy-level diagram for the ESSEM caused by the ground-state mixing. (c) Illustration of the magnetic field orientation. (d) Energies of the ground states ($E_\pm$) and the energy splitting ($\delta$) as functions of the magnetic field. The solid lines are the exact diagonalization with the full Hamiltonian in Eq. (S12) (including the strains $\Pi_x = 2.2\ MHz$ and $\Pi_z = 1.1\ MHz$) and the dashed lines are the results from Eq. (S16, S17).

Figure S5 is the ESEEM signals under different microwave power excitations with a known longitudinal magnetic field ($B_z = 0.01$ mT). Under weak microwave excitation (with Rabi frequency $\Omega = 1.25$ MHz and $\Omega < \Delta$), only a very slow modulation is observed. When increasing the excitation power to $\Omega = 5$ MHz, transitions involving different $^{15}$N states can be excited by the microwave pulse, and the ESEEM modulation from the host $^{15}$N nuclear spin is observed (evidenced by the characteristic ~ 3 MHz coupling in the FFT curve).

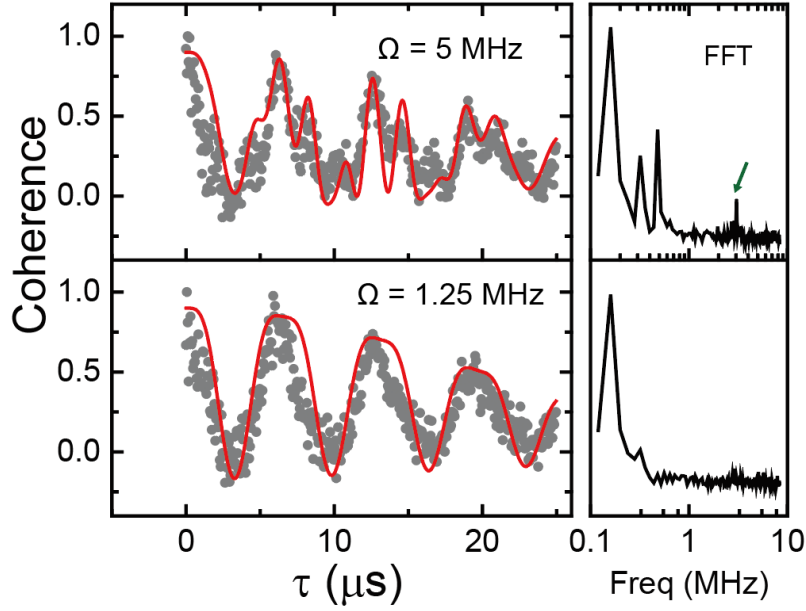

**Figure S5.** Electron spin echo envelop modulation (ESEEM) signal at $B_z = 0.01\ mT$ under different microwave power excitations and the FFTs of the ESEEM signals. The green arrow indicates the characteristic frequency due to the coupling to the host $^{15}$N.

We compare the calculated the ESSEM signal with the measured one. From Eq. (S17), we find that the azimuth angle ($\phi$) of the magnetic field only contributes a small modulation of the ground-state splitting ($\delta$). For simplicity, we assume the transverse magnetic field is in the $xz$-plane defined by the local strains. The Hamiltonian in Eq. (S12) in the reference frame rotating with the microwave frequency becomes

$$H_{\mathrm{RW}} = (D_{\mathrm{gs}} + \Pi_z - \omega)S_z^2 + \Pi_x(S_xS_x - S_yS_y) + \gamma_e B_z S_z + \gamma_e B_x S_x + \mathbf{S} \cdot \mathbb{A}_C \cdot \mathbf{I}_C + \Omega S_x, \quad (S18)$$

where $\omega$ is the frequency of the microwave and $\Omega$ is the Rabi frequency of the microwave pulses. The initial electron spin state is polarized to the $m_s = 0$ state and the nuclear spin is in a mixed state with a density matrix $\rho_N = 1/2$. The initial density matrix of the system is

$$\rho_0 = |0\rangle\langle 0| \otimes \rho_N. \quad (S19)$$

In our simulation, we consider the subspace expanded by the states $|m_s = 0, \mu_C = +1/2\rangle$, $|m_s = 0, \mu_C = -1/2\rangle$, and $|m_s = -1, \mu_C = +1/2\rangle$ for the resonance frequency of the microwave pulses. The evolution operator $U_\tau$ for a spin echo sequence for an evolution time $\tau$ is

$$U_\tau = R_{\frac{\pi}{2}} U\left(\frac{\tau}{2}\right) R_\pi U\left(\frac{\tau}{2}\right) R_{\frac{\pi}{2}} = e^{-\frac{iH_{\mathrm{RW}}t_\pi}{2}} e^{-\frac{iH'_{\mathrm{RW}}\tau}{2}} e^{-iH_{\mathrm{RW}}t_\pi} e^{-\frac{iH'_{\mathrm{RW}}\tau}{2}} e^{-\frac{iH_{\mathrm{RW}}t_\pi}{2}}. \quad (S20)$$

Here $H'_{RW}$ is the Hamiltonian correspond to the free evolution where $\Omega$ is zero. The echo signal is

$$S(\tau) = \sum_{\mu_C = \pm\frac{1}{2}} \langle 0, \mu_C | U_\tau \rho_0 U_\tau^\dagger | 0, \mu_C \rangle. \tag{S21}$$

We calculate the spin echo signal numerically with local strains $\Pi_x = 2.2$ MHz and $\Pi_z = 1.1$ MHz. For $B_z = 0.01$ mT, the calculated echo signals with different $B_x$ are shown in Fig. S6 (a). The very slow modulation frequency changes with the magnetic field $B_x$. In Fig. S6 (b), we plot the modulation frequencies as a function of the energy splitting $\delta$. The modulation frequency changes linearly with the ground-state energy splitting $\delta$. Using Eq. (S17), we can estimate the magnitude of the transverse magnetic field. In our experiment, the modulation frequency of 0.13 MHz in Fig. S5 corresponds to a transverse magnetic field about 0.03 mT.

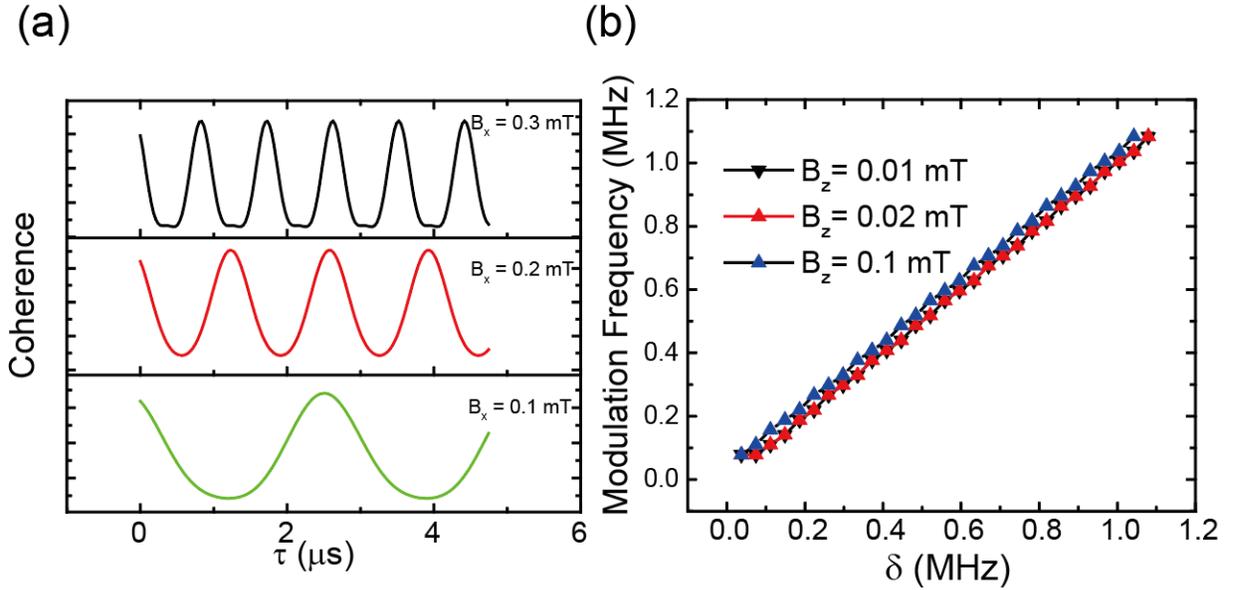

**Figure S6.** Calculated ESEEM under different magnetic fields. (a) ESEEM signals under different magnetic fields with a fixed $B_z = 0.01\ mT$. (b) The modulation frequencies as a function of the ground-state energy splitting.

Note 4.2 Modulation induced by weakly coupled nuclear spins

Here we consider the coupling to nuclear spins located far away from the NV center, which have very weak hyperfine couplings. The ESEEM signal is well studied before [37, 40-42]. Here we regard the NV electron spin and a first-shell $^{13}$C nuclear spin as a hybrid spin **S** and analyze its coupling to the remote nuclear spins. The total Hamiltonian becomes Eq. (S18) with an additional perturbation $H_V$ as

$$H = H_{\text{NV-C}} + H_V, \tag{S22a}$$

$$H_V = \sum_i \mathbf{S} \cdot \mathbb{A}_i \cdot \mathbf{I}_i, \tag{S22b}$$

where $\mathbb{A}_i$ is the hyperfine tensor of the $i$th nuclear spin $\mathbf{I}_i$. In the eigenstate basis of the hybrid central spin $\{|\alpha\rangle\}$, the off-diagonal coupling of $H_V$ can be dropped since the hyperfine tensors $\mathbb{A}_i$ are much weaker than the energy splitting of the central spin $\mathbf{S}$, and the diagonal terms are

$$\langle \alpha | H_V | \alpha \rangle = \sum_i \langle \alpha | \mathbf{S} | \alpha \rangle \cdot \mathbb{A}_i \cdot \mathbf{I}_i \equiv \sum_i \mathbf{A}_{i,\alpha} \cdot \mathbf{I}_i. \tag{S23}$$

The effective field acts on each nuclear spin for different eigenstate of the central hybrid spin is

$$\mathbf{h}_i^\alpha = \mathbf{B} - \frac{\mathbf{A}_{i,\alpha}}{\gamma_n}. \tag{S24}$$

The final ESEEM signal of the transition between $|\alpha\rangle$ and $|\beta\rangle$ is

$$P(\tau) = 1 - \frac{2\left|\mathbf{h}_i^\alpha \times \mathbf{h}_i^\beta\right|^2}{\left|\mathbf{h}_i^\alpha\right|^2 \left|\mathbf{h}_i^\beta\right|^2} \sin^2 \frac{|\gamma_n \mathbf{h}_i^\alpha| \tau}{2} \sin^2 \frac{|\gamma_n \mathbf{h}_i^\beta| \tau}{2}. \tag{S25}$$

In particular, when one of the eigenstates, e.g., $|\alpha\rangle$ is the state with $m_s = 0$, the hyperfine coupling is nearly zero ($A_{i,\alpha} \approx 0$) and all the nuclear spins have $\mathbf{h}_i^\alpha = \mathbf{B}$ and have the same echo recovery time $\tau = \frac{2n\pi}{\gamma_n B}$ for an integer number $n$. Physically, this is because the evolution of all the nuclear spins during such a period when the central spin in state $|\alpha\rangle$ amounts to a null evolution.

## Note 5. Pulse sequences for relaxometry

The relaxation dynamics of an NV center spin strongly coupled to a $^{13}$C nuclear spin is illustrated in Fig. S7 (a). Since the relaxation time of the nuclear spin is much longer than that of the electron spin, the nuclear spin is conserved during the relaxation of the electron spin, resulting in two independent processes [left and right panels in Fig. S7 (a)]. The relaxation dynamics is reduced to a three-level system with the rate equation as

$$\frac{d}{dt}\begin{pmatrix} P_{0,\mp 1/2} \\ P_{+1,\mp 1/2} \\ P_{-1,\mp 1/2} \end{pmatrix} = \begin{pmatrix} -\Gamma_{1/3} - \Gamma_{2/4} & \Gamma_{1/3} & \Gamma_{2/4} \\ \Gamma_{1/3} & -\Gamma_{1/3} - \gamma_{\alpha/\beta} & \gamma_{\alpha/\beta} \\ \Gamma_{2/4} & \gamma_{\alpha/\beta} & -\Gamma_{2/4} - \gamma_{\alpha/\beta} \end{pmatrix} \begin{pmatrix} P_{0,\mp 1/2} \\ P_{+1,\mp 1/2} \\ P_{-1,\mp 1/2} \end{pmatrix}, \tag{S26}$$

where $P_{m_s,\mu_C}$ is the population of the corresponding spin state, and $\Gamma_i$ and $\gamma_{\alpha/\beta}$ are the relaxations rates for different transitions shown in Fig. S7 (a). When there is a magnetic nanoparticle in the proximity, the main mechanism of the spin relaxation is the magnetic noise from the nanoparticle, which has little effect on the transition between $|m_s = +1\rangle$ and $|m_s = -1\rangle$, characterized by the rates $\gamma_{\alpha/\beta}$.

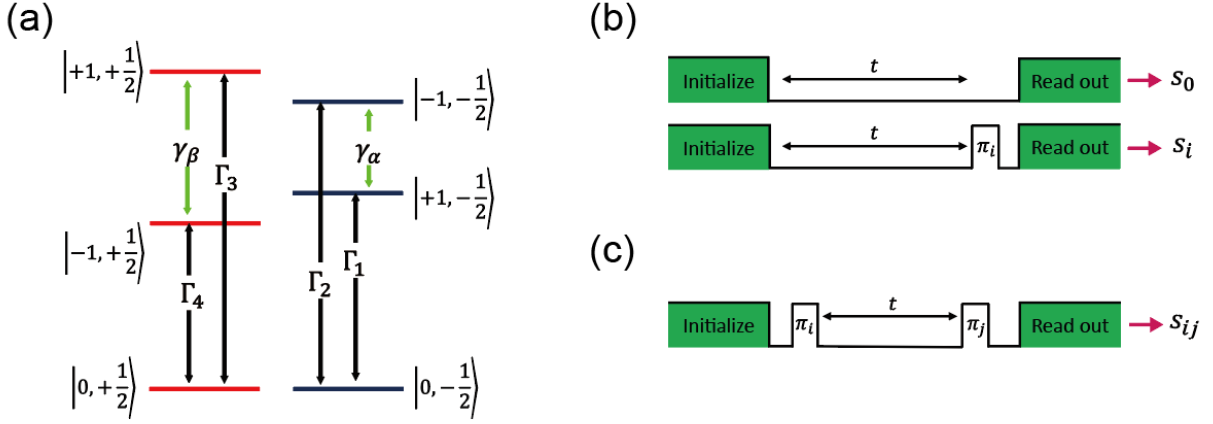

**Figure S7.** Relaxation dynamics and pulse sequences for $T_1$ measurement. (a) Relaxation rates between different states. The transition rate ($\gamma_{\alpha/\beta}$) between $|m_s = +1\rangle$ and $|m_s = -1\rangle$ is much smaller than $\Gamma_i$ when the noise is mainly from the magnetic nanoparticle. (b) Pulse sequences to measure the $T_1$ time. (c) Pulse sequences to measure double quantum relaxations.

Fig. S7 (b) shows the pulse sequences for $T_1$ measurement. To exclude the effects of fluorescence decay due to spin-independent processes (such as the charge state jumps), we use the difference between the signals with and without a π pulse between the states $|0, \mu_C\rangle$ and $|\pm 1, \mu_C\rangle$ right before readout. We first polarize the spin state to the states $|m_s = 0, \mu_C = \pm 1/2\rangle$ and let the system relax for a time $t$. At the end of the sequence, we measure the spin-dependent fluorescence either directly or after applying a π pulse on the transition. The signals are denoted as $s_0$ (without a π pulse) and $s_i$ (with a π pulse applied to the $i$-th transition with the relaxation rate $\Gamma_i$). The signals are related to the populations in different states by

$$s_0 = A\left(P_{0,-\frac{1}{2}} + P_{0,+\frac{1}{2}}\right) + B\left(P_{-1,-\frac{1}{2}} + P_{+1,-\frac{1}{2}} + P_{-1,+\frac{1}{2}} + P_{+1,+\frac{1}{2}}\right) + C, \quad (S27a)$$

$$s_1 = A\left(P_{+1,-\frac{1}{2}} + P_{0,+\frac{1}{2}}\right) + B\left(P_{-1,-\frac{1}{2}} + P_{0,-\frac{1}{2}} + P_{-1,+\frac{1}{2}} + P_{+1,+\frac{1}{2}}\right) + C, \quad (S27b)$$

$$s_2 = A\left(P_{-1,-\frac{1}{2}} + P_{0,+\frac{1}{2}}\right) + B\left(P_{0,-\frac{1}{2}} + P_{+1,-\frac{1}{2}} + P_{-1,+\frac{1}{2}} + P_{+1,+\frac{1}{2}}\right) + C, \quad (S27c)$$

$$s_3 = A\left(P_{0,-\frac{1}{2}} + P_{+1,+\frac{1}{2}}\right) + B\left(P_{-1,-\frac{1}{2}} + P_{+1,-\frac{1}{2}} + P_{-1,+\frac{1}{2}} + P_{0,+\frac{1}{2}}\right) + C, \quad (S27d)$$

$$s_4 = A\left(P_{0,-\frac{1}{2}} + P_{-1,+\frac{1}{2}}\right) + B\left(P_{-1,-\frac{1}{2}} + P_{+1,-\frac{1}{2}} + P_{0,+\frac{1}{2}} + P_{+1,+\frac{1}{2}}\right) + C, \quad (S27e)$$

where $A$ and $B$ are the photon counts for the $|m_s = 0\rangle$ and $|m_s = \pm 1\rangle$ states respectively, and $C$ is for the background photon counts. With the relation $P_{0,\mu_C} + P_{+1,\mu_C} + P_{-1,\mu_C} = 1/2$, we obtain the populations of each spin state from the photon counts as

$$P_{0,\mp\frac{1}{2}} = \frac{1}{6} + \frac{2s_0 - (s_{1/3} + s_{2/4})}{3(A-B)}, \tag{S28a}$$

$$P_{+1,\mp\frac{1}{2}} = \frac{1}{6} + \frac{2s_{1/3} - (s_0 + s_{2/4})}{3(A-B)}, \tag{S28b}$$

$$P_{-1,\mp\frac{1}{2}} = \frac{1}{6} + \frac{2s_{2/4} - (s_0 + s_{1/3})}{3(A-B)}. \tag{S28c}$$

For signals more sensitive to the transition rates $\gamma_{\alpha/\beta}$, we measure the double quantum (DQ) relaxation [27]. Fig. S7 (c) is the sequence for measuring the double quantum relaxation between $|m_s = +1; \mu_C\rangle$ and $|m_s = -1; \mu_C\rangle$. We first prepare the states to $|m_s = +1; \mu_C\rangle$ and let the system evolve a time $t$. At the end of the sequence, an additional $\pi$ pulse is applied on the transition $|m_s = 0; \mu_C\rangle \leftrightarrow |m_s = +1; \mu_C\rangle$ or $|m_s = 0; \mu_C\rangle \leftrightarrow |m_s = -1; \mu_C\rangle$ before readout. The signal $s_{ij}$ is for the two $\pi$ pulses applied for the $i$-th and $j$-th transitions. The measured signals are related to the populations in different states as

$$s_{11} = A\left(P'_{+1,-\frac{1}{2}} + P'_{0,+\frac{1}{2}}\right) + B\left(P'_{-1,-\frac{1}{2}} + P'_{0,-\frac{1}{2}} + P'_{-1,+\frac{1}{2}} + P'_{+1,+\frac{1}{2}}\right) + C, \tag{S29a}$$

$$s_{12} = A\left(P'_{-1,-\frac{1}{2}} + P'_{0,+\frac{1}{2}}\right) + B\left(P'_{0,-\frac{1}{2}} + P'_{+1,-\frac{1}{2}} + P'_{-1,+\frac{1}{2}} + P'_{+1,+\frac{1}{2}}\right) + C, \tag{S29b}$$

$$s_{33} = A\left(P'_{0,-\frac{1}{2}} + P'_{+1,+\frac{1}{2}}\right) + B\left(P'_{-1,-\frac{1}{2}} + P'_{+1,-\frac{1}{2}} + P'_{-1,+\frac{1}{2}} + P'_{0,+\frac{1}{2}}\right) + C, \tag{S29c}$$

$$s_{34} = A\left(P'_{0,-\frac{1}{2}} + P'_{-1,+\frac{1}{2}}\right) + B\left(P'_{-1,-\frac{1}{2}} + P'_{+1,-\frac{1}{2}} + P'_{0,+\frac{1}{2}} + P'_{+1,+\frac{1}{2}}\right) + C. \tag{S29d}$$

Here $P'_{m_s,\mu_C}$ differ from $P_{m_s,\mu_C}$ for their different initial conditions. The DQ relaxations between $|m_s = +1; \mu_C\rangle$ and $|m_s = -1; \mu_C\rangle$ are

$$S_{DQ,-\frac{1}{2}} = P'_{+1,-\frac{1}{2}} - P'_{-1,-\frac{1}{2}} = \frac{s_{11} - s_{12}}{A - B}, \tag{S30a}$$

$$S_{DQ,+\frac{1}{2}} = P'_{+1,+\frac{1}{2}} - P'_{-1,+\frac{1}{2}} = \frac{s_{33} - s_{34}}{A - B}, \tag{S30b}$$

The measured populations and DQ relaxation signals of NV3-13C with an MNP nearby are shown in Fig. S8. The upper panels of Fig. S8 (a, b) show the populations for $|m_s = 0, \pm 1, \mu_C\rangle$ obtained with Eq. (S28) and the lower panels show the DQ relaxations between $|m_s = +1, \mu_C\rangle$ and $|m_s = -1, \mu_C\rangle$ obtained with Eq. (S30). We fit the populations in $|m_s, \mu_C = -1/2\rangle$ and the DQ signal for $\mu_C = -1/2$ [curves in Fig. S8 (a)] by solving the rate equations in Eq. (S26) with their respective initial state, obtaining the best fitted relaxation rates $\Gamma_1 = 10.0(5)$ ms$^{-1}$, $\Gamma_2 = 5.0(4)$ ms$^{-1}$, and $\gamma_\alpha = 0.5(2)$ ms$^{-1}$. Similarly, we fit the curves in Fig. S8 (b) and get the relaxation rates $\Gamma_3 = 6.0(4)$ ms$^{-1}$, $\Gamma_4 = 8.9(5)$ ms$^{-1}$, and $\gamma_\beta = 0.5(2)$ ms$^{-1}$. The measured relaxation rates $\Gamma_i$ increase dramatically as compared with the relaxation rate of NV3-C13 [0.15(1) ms$^{-1}$] before inducing the MNP, which indicates that the magnetic noise is mainly from the nearby MNP. The relaxation rates $\Gamma_1$ and $\Gamma_4$ for the transitions with lower energies are significantly larger than $\Gamma_2$ and $\Gamma_3$ for the transitions with higher energies, and the transitions with similar frequencies have similar relaxation rates. These facts suggest that the noise from the magnetic nanoparticle has a sub-GHz spectral width. The DQ relaxation rates are nearly the same ($\gamma_\alpha = \gamma_\beta$) and are much smaller than the single-quantum relaxation rates $\Gamma_i$, which is consistent with the results in Ref. [27].

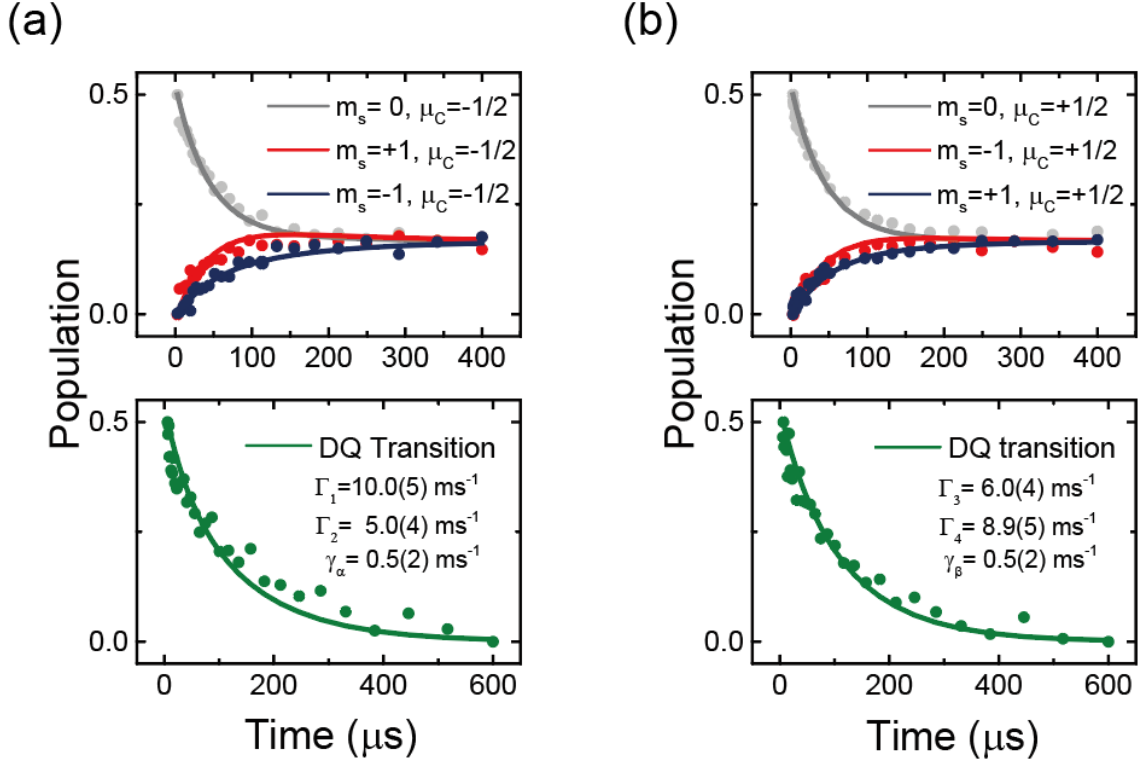

**Figure S8.** Populations of the spin states and the double-quantum signals between $|m_s = +1; \mu_C\rangle$ and $|m_s = -1; \mu_C\rangle$. (a) Upper: populations of the states $|m_s; \mu_C = -1/2\rangle$; lower: DQ relaxation between

$|m_s = +1, \mu_C = -1/2\rangle$ and $|m_s = -1, \mu_C = -1/2\rangle$. (b) Upper: populations of the states $|m_s; \mu_C = +1/2\rangle$; lower: DQ relaxation between $|m_s = +1, \mu_C = +1/2\rangle$ and $|m_s = -1, \mu_C = +1/2\rangle$. The dots are measured spin populations and DQ relaxations and the curves are fitting results from Eq. (S26).